\documentclass[12pt]{article}
\usepackage{graphicx}

\usepackage{mathtools}
\usepackage{subcaption}

\newcommand\pubnumber{NuPhys2016-Arnold}
\newcommand\pubdate{April 2017}

\def\napoli{University of Bristol\\School of Physics\\
United Kingdom}
\def\support{\footnote{for the SoLid collaboration}}

\def\Title#1{\begin{center} {\Large #1 } \end{center}}
\def\Author#1{\begin{center}{ \sc #1} \end{center}}
\def\Address#1{\begin{center}{ \it #1} \end{center}}

\newcommand\pubblock{\rightline{\begin{tabular}{l} \pubnumber\\
         \pubdate  \end{tabular}}}
\newenvironment{Abstract}{\begin{quotation}  }{\end{quotation}}
\newenvironment{Presented}{\begin{quotation} \begin{center} 
             PRESENTED AT\end{center}\bigskip 
      \begin{center}\begin{large}}{\end{large}\end{center} \end{quotation}}
\def\Acknowledgements{\bigskip  \bigskip \begin{center} \begin{large}
             \bf ACKNOWLEDGEMENTS \end{large}\end{center}}
\newcommand{\LiZnS}{$\prescript{6}{}{\mathrm{LiF:ZnS(Ag)}}$}

\def\beq{\begin{equation}}
\def\eeq#1{\label{#1}\end{equation}}
\def\eeqn{\end{equation}}

\def\beqa{\begin{eqnarray}}
\def\eeqa#1{\label{#1}\end{eqnarray}}
\def\eeqan{\end{eqnarray}}

\let\bar=\overbar

\def\Dslash{\not{\hbox{\kern-4pt $D$}}}
\def\dslash{\not{\hbox{\kern-2pt $\del$}}}

\def\msb{{\bar{\ssstyle M \kern -1pt S}}}

\begin{document}
\begin{titlepage}
\pubblock

\vfill
\Title{Trigger for the SoLid Reactor Antineutrino Experiment}
\vfill
\Author{Lukas On Arnold\support}
\Address{\napoli}
\vfill
\begin{Abstract}
SoLid, located at SCK-CEN in Mol, Belgium, is a reactor antineutrino experiment at a very short baseline of 5.5 -- 10m aiming at the search for sterile neutrinos and for high precision measurement of the neutrino energy spectrum of Uranium-235. It uses a novel approach using Lithium-6 sheets and PVT cubes as scintillators for tagging the Inverse Beta-Decay products (neutron and positron).  
Being located overground and close to the BR2 research reactor, the experiment faces a large amount of backgrounds.  Efficient real-time background and noise rejection is essential in order to increase the signal-background ratio for precise oscillation measurement and decrease data production to a rate which can be handled by the online software. Therefore, a reliable distinction between the neutrons and background signals is crucial. This can be performed online with a dedicated firmware trigger. A peak counting algorithm and an algorithm measuring time over threshold have been identified as performing well both in terms of efficiency and fake rate, and have been implemented onto FPGA.

\end{Abstract}
\vfill
\begin{Presented}
NuPhys2016, Prospects in Neutrino Physics\\
Barbican Centre, London, UK,  December 12--14, 2016
\end{Presented}
\vfill
\end{titlepage}
\def\thefootnote{\fnsymbol{footnote}}
\setcounter{footnote}{0}

\section{Introduction}
SoLid is a short-baseline reactor antineutrino experiment. It probes the deficit from the electron antineutrino ($\bar{\nu}_e$) prediction yield measured by various reactor antineutrino expriments -- known as the reactor antineutrino anomaly \cite{Mention:2011rk} -- in close proximity ($5.5\mathrm{m}\ldots10\mathrm{m}$) from the reactor core.  An observation of an oscillation in the antineutrino energy spectrum at close distance measured independentdly of the predictions, relative measurements at different distances, could provide evidence for the existence of sterile neutrinos, the sterile flavour state being an addition flavour to the three flavours commonly known ($\nu_e$, $\nu_\mu$ and $\nu_\tau$) that does not interact weakly.

\begin{figure}[htb]
\centering
\begin{subfigure}{.5\textwidth}
	\centering
	\includegraphics[trim={6cm 0 0 0},clip,height=2.0in]{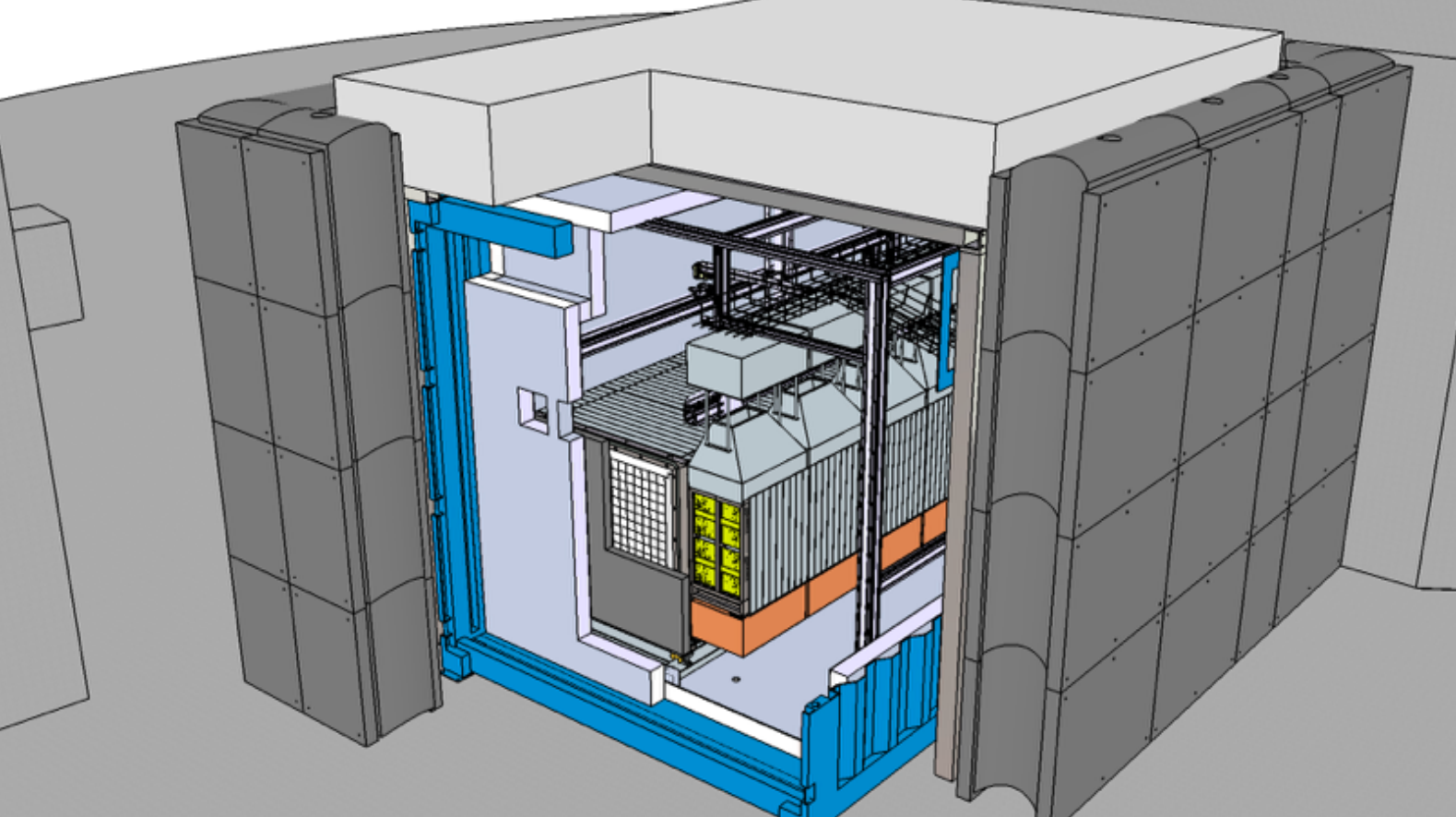}
	\caption{The SoLid $1.6\mathrm{t}$-detector, consisting of 50 planes.}
	\label{fig:solid}
\end{subfigure}
\begin{subfigure}{.4\textwidth}
	\centering
	\includegraphics[trim={0 0 31cm 0},clip,height=2.5in]{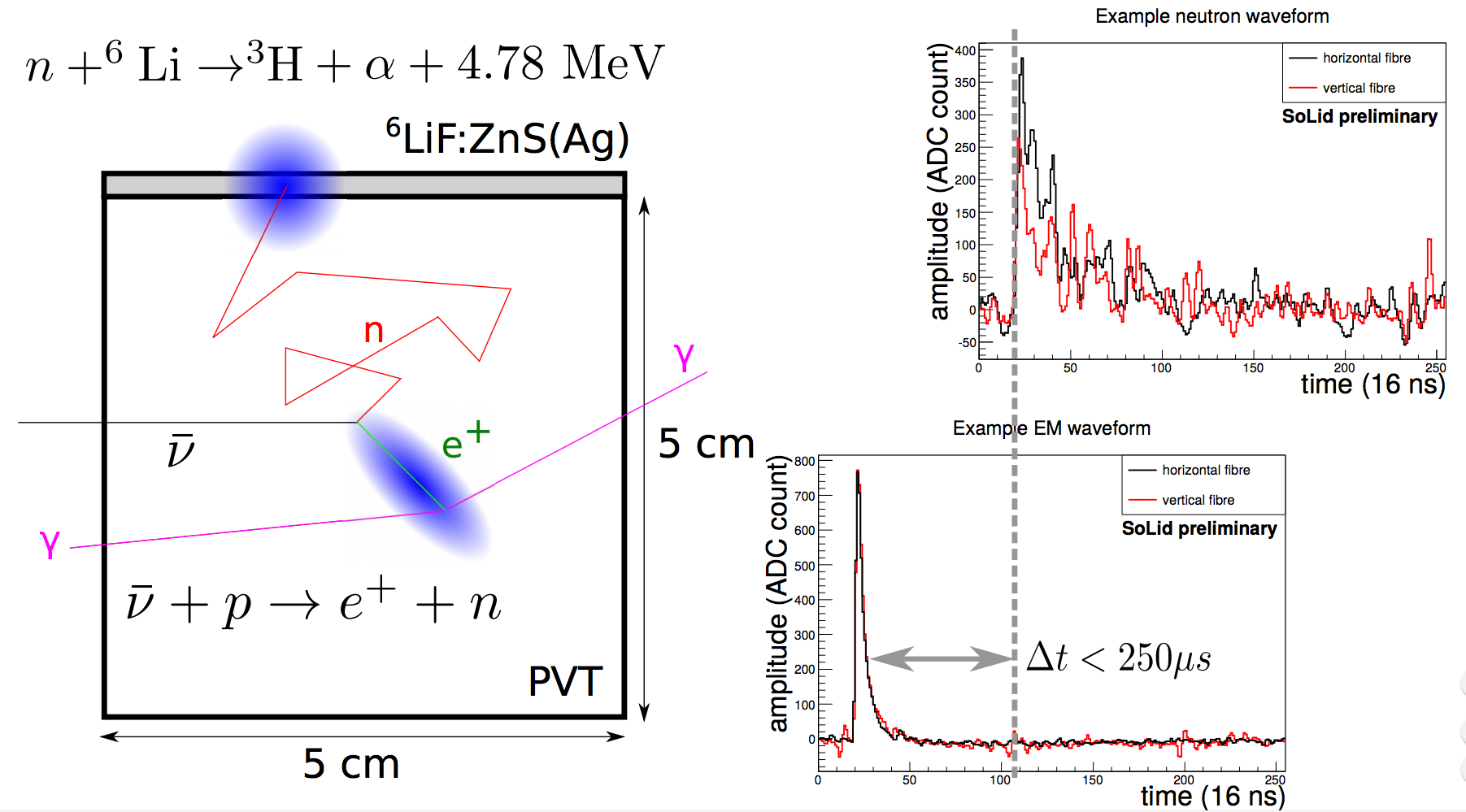}
	\caption{Interaction of $n$ and $e^+$ in the detector cube. \textit{Source:} \cite{Ryder:2015sma}.}
	\label{fig:catch}
\end{subfigure}
\caption{}
\end{figure}

The experiment is located at the BR2 reactor at SCK\textbullet CEN in Mol, Belgium. BR2 is a tank-type material research reactor fuelled by Uranium-235 with a power range of up to $100\mathrm{MW}$ and a core diameter of only $50\mathrm{cm}$, which makes it an almost point-like neutrino source. The SoLid detector, depicted in Figure~\ref{fig:solid}, consists of 5 modules, or 50 planes that each contain 256 $(5\times5\times5)\mathrm{cm}$ cubes consisting of PVT and a layer of \LiZnS. The detector uses $1.6\mathrm{t}$ of active material.
\subsection{Detector Technology}
SoLid's neutrino detection is based on Inverse Beta-Decay (IBD):
\begin{equation}
\bar{\nu}_e+p\rightarrow e^+ + n.
\end{equation}
The two IBD  products -- the positron ($e^+$) and the neutron ($n$) -- are captured by two different materials in the cube. The positron is absorbed within $\mathcal{O}\left({10^{-8}}\right)\mathrm{s}$ by the PVT and is emitted as a short ($\mathcal{O}\left({10^{-7}}\right)\mathrm{s}$), intense light pulse \cite{Ryder:2015sma}\cite{Michiels:2016qui} . 
The neutron however undergoes thermalisation while being scattered through the material before being caught by the \LiZnS\phantom{a}sheets by the process
\begin{equation}
n+\prescript{6}{}{\mathrm{Li}}\rightarrow \prescript{3}{}{\mathrm{H}}+\alpha+4.78\mathrm{MeV}.
\end{equation}

Both $\prescript{3}{}{\mathrm{H}}$ and the $\alpha$-particle contain sufficient energy to excite the electrons in the $\mathrm{ZnS}$ crystals. Scintillation light is emitted by de-excitation of these electrons over a longer period of $\mathcal{O}\left(10^{-6}\right)\mathrm{s}$, an order of magnitude higher than the positron's signal length. The positron and neutron signals hence are clearly distinct by their latency, amplitude and decay time. IBD capture by the cube is sketched in Figure~\ref{fig:catch}.

\subsection{Read-out System}
Wavelength-shifting fibres guide the light from the PVT cubes to Silicon Photomultipliers (SiPMs). 64 SiPMs are read out per plane. The SiPM signal are shaped and filtered by an analogue board and then sent to the digital board that includes the FPGA. The read-out electronics is placed on side of the frame containing the cubes. The boards are custom-made to serve the needs of the SoLid experiment. \cite{Arnold:2017lph}

\subsection{Firmware}
The SoLid FPGA firmware is responsible for buffering the data, triggering on it, and for communication with other planes as well as with the data acquisition device. Also, it has the slow-control for the SiPMs integrated in its functionality. The firmware is based on the \textit{IPbus} protocol, a gigabit Ethernet-based reliable high-performance protocol designed for particle physics experiments \cite{Larrea:2015wra} \cite{Dasgupta:2015ohj}. 
A medium-density device -- the \textsc{Xilinx} Artix 7-200T FPGA -- is used.
The ionising radiation being faced is about the same as normal background levels as the increase in neutron and gamma flux by the reactor is compensated for by water and containment shielding.

\section{Trigger}
SoLid faces a large amount of background signals due to it being an overground experiment in proximity to the reactor core. This requires the read-out system -- or namely the trigger part  -- to reject as many background signals as possible. 
The extent to which this rejection has to take place is dependent on data generation by the digital board and on the data handling capacity by the online software.
A summary of data rates and reduction factors is given in Table~\ref{tab:datarates}. It is the task of the trigger to provide efficient, yet pure data reduction.
\begin{table}[t]
\begin{center}
\begin{tabular}{l|ccc}  
\textbf{Stage}  & \textbf{Data rate $[\mathrm{s}^{-1}]$} & \textbf{Data rate $[\mathrm{d}^{-1}]$} & \textbf{Reduction} \\ \hline
Digital board   &    $=1.8\mathrm{Tb}$                &             $=19\mathrm{PB}$                           &                         --      \\
Online software (maximum) &   $\sim1.0\mathrm{Gb}$                 & $\sim11\mathrm{TB}$                   &                          $\sim1800$     \\
Data storage &  $\sim10\mathrm{Mb}$                                        &$\sim100\mathrm{GB}$                                &     $\sim100$ \\ \hline
\end{tabular}
\caption{Data rates at the different stages of the data acquisition chain per second and per day. Reduction factor is given to the previous stage.}
\label{tab:datarates}
\end{center}
\end{table}
The positron signal -- due to its high amplitude and briefness -- is fairly easy to distinguish; a threshold trigger is used. In contrast, the neutron trigger is more complex, as its signals usually do not reach high amplitudes. Using solely the positron trigger -- that has low purity -- would increase data rate up to a point at which it cannot be handled by the online software, which is the reason why triggering properly on the neutrons is necessary for detector read-out.

\subsection{Algorithm Evaluation}
Test data have been used for evaluation of different neutron trigger algorithms, or features. The test setup uses an $\prescript{241}{}{\mathrm{AmBe}}$ $\alpha$-particle source, located $\sim 3\mathrm{cm}$ from a PVT cube connected to two SiPMs and a photomultiplier tube that is used for acquiring the reference signal.

\begin{figure}[htb]
\centering
	\includegraphics[height=3.5in]{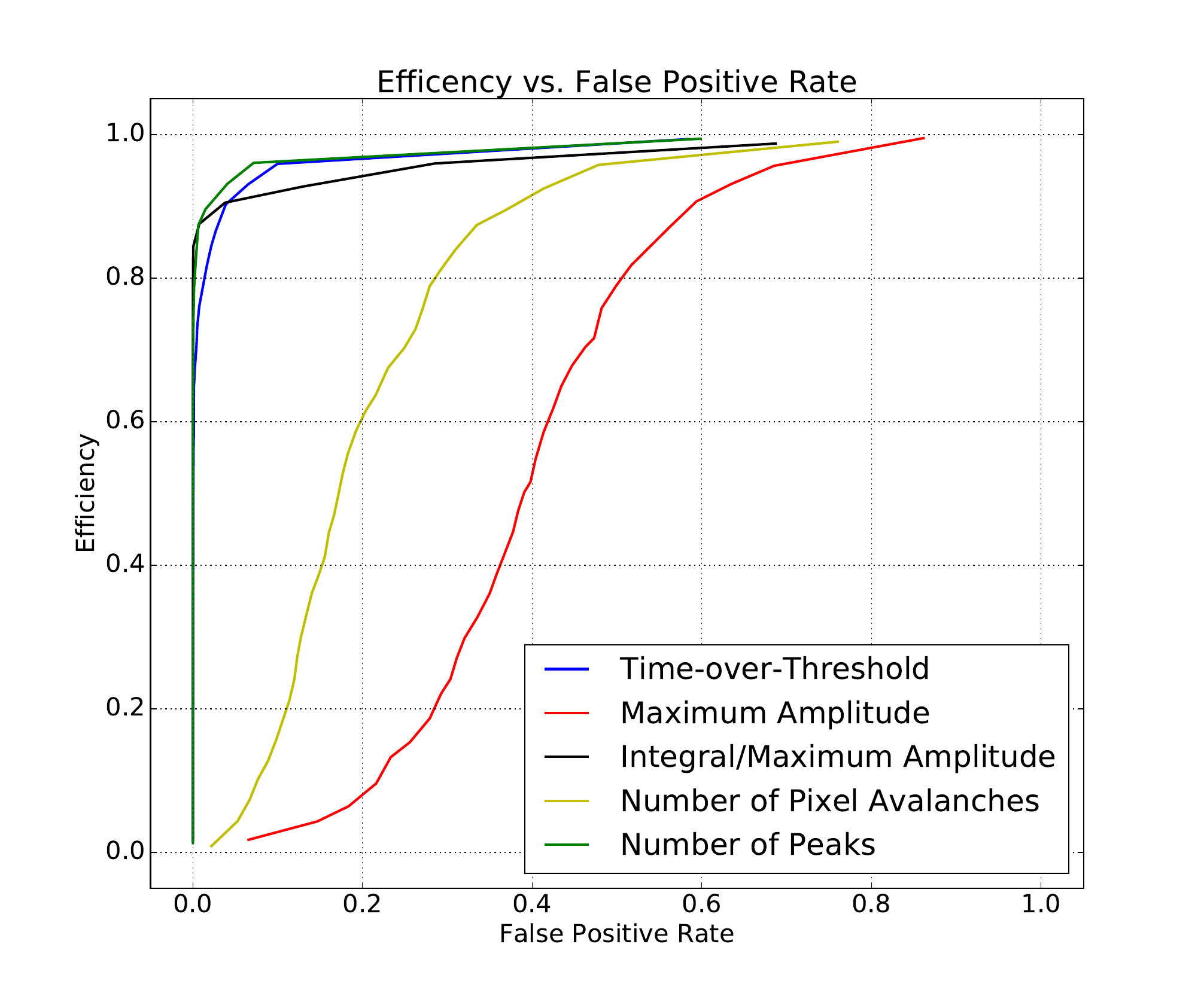}
	\caption{Summarising ROC curve of different features in terms of efficiency and fake rate ($=$ false positive rate), with the Number-of-Peaks algorithm performing best on the test data set.}
	\label{fig:comp}
\end{figure}

 An overall comparison of features has been used to reduce the number of potential features by short-listing those that perform well, i.e. that yield high efficiency, high purity and low fake rate simultaneously. 
High efficiency, or true positive rate, correlates with how many IBD physics events are caught. High purity, or positive predictive value, correlates with contamination of IBD events with backgrounds. And a low fake rate, or false positive rate,  means that few non-IBD events are acquired. 
The results on the efficiency-fake rate plane are shown in figure~\ref{fig:comp}.
\begin{figure}[htb]
\centering
\begin{subfigure}{.5\textwidth}
  \centering
  \includegraphics[width= \linewidth]{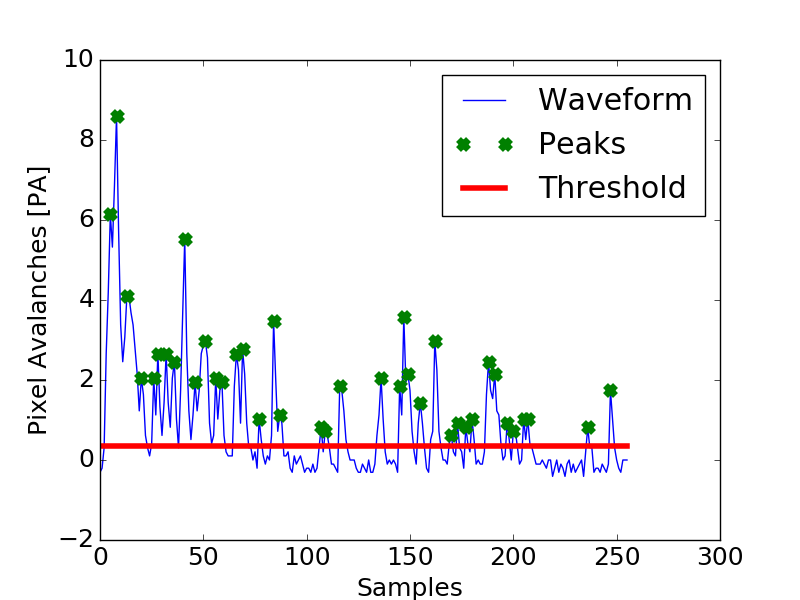}
  \caption{Neutron signal.}
  \label{fig:sub10}
\end{subfigure}%
\begin{subfigure}{.5\textwidth}
  \centering
  \includegraphics[width= \linewidth]{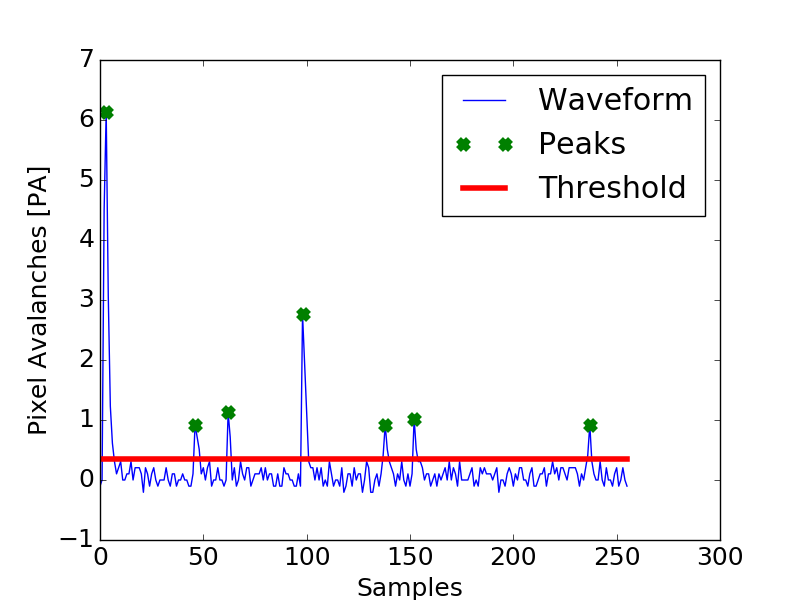}
  \caption{Dark count signal.}
  \label{fig:sub20}
\end{subfigure}
\caption{Peak finding algorithm. Green crosses represent found maxima above threshold.}
\label{fig:npeak}
\end{figure}

A variety of trigger algorithms have been evaluated. Of these, the following three have been performing well on test data:

\paragraph{Number-of-Peaks}
The Number-of-Peaks feature counts the number of maxima above a certain threshold $\theta$ within a time window. An example is shown in figure~\ref{fig:npeak}. The threshold is set in order that pure noise-induced maxima are not considered. The reasoning behind this algorithm is that each time the \LiZnS-layer is emitting a photon, a peak on the input signal occurs.  It may be expressed in discrete space as a function of the discrete signal within a limited time window $X$:
\begin{equation}
\label{eq:npeaksd2}
g\left(X\right)=\left\vert \left\{ t| \forall t:{X[t]>\theta} \land{ X[t-1]\geq X[t-2] \land X[t]<X[t-1]}\right\}\right\vert.
\end{equation}
In other words, the cardinality -- i.e. the number of elements -- $\left|.\right|$ of the set $\{.\}$ of points that are a maximum above a certain threshold forms the feature.
\paragraph{Time-over-Threshold}
The Time-over-Threshold measures the number of samples (i.e. length of time) a signal is above a certain threshold $\theta$. It is expressed as a a function of a discrete signal $X$ within a time window of sample length $m$ as
\begin{equation}
\label{eq:totd}
g\left(X\right)=
{
\sum_{t=1}^{m}\delta[t]} \land {\delta[t] = \begin{cases}
  1, & \text{if } X[t] > \theta, \\
  0, & \text{otherwise}.
\end{cases}}
\end{equation}

\paragraph{Integral-over-Amplitude}
The Integral-over-Amplitude (or Integral$/$maximum Amplitude) divides the integral of a signal by division by maximum amplitude:
\begin{equation}
\label{eq:integralamplitude2}
g\left(X\right)=\frac
{
\sum_{t=1}^{m} X[t]}{\max X}.
\end{equation}

\subsection{Feature Tuning}
\begin{figure}[htb]
\centering
\begin{subfigure}{.47\textwidth}
  \centering
  \includegraphics[width= \linewidth]{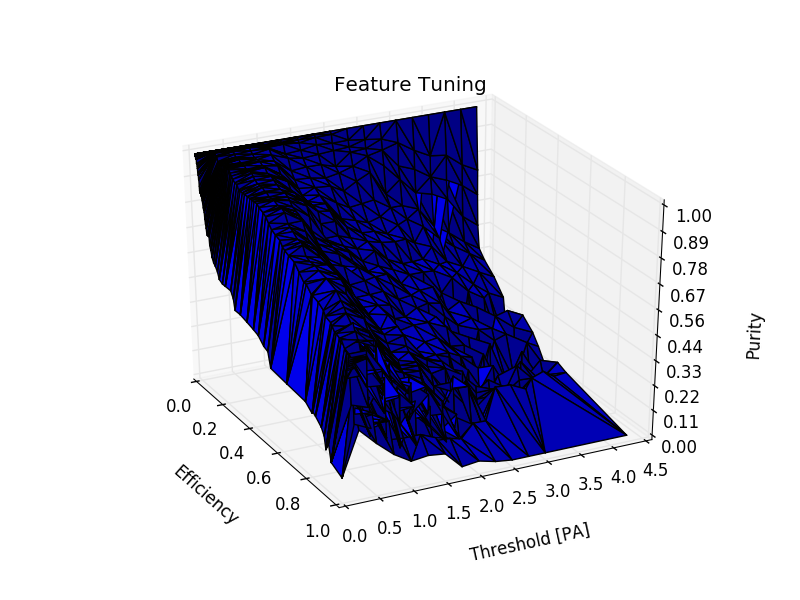}
  \caption{Surface plot for Time-over-Threshold feature tuning. Threshold is swept from $0$ to $4.5\mathrm{PA}$ with a mountain visible in the range of $0.5\mathrm{PA}$, indicating the optimum value.}
  \label{fig:tottuning}
\end{subfigure}%
\hspace{5mm}
\begin{subfigure}{.47\textwidth}
  \centering
  \includegraphics[width= \linewidth]{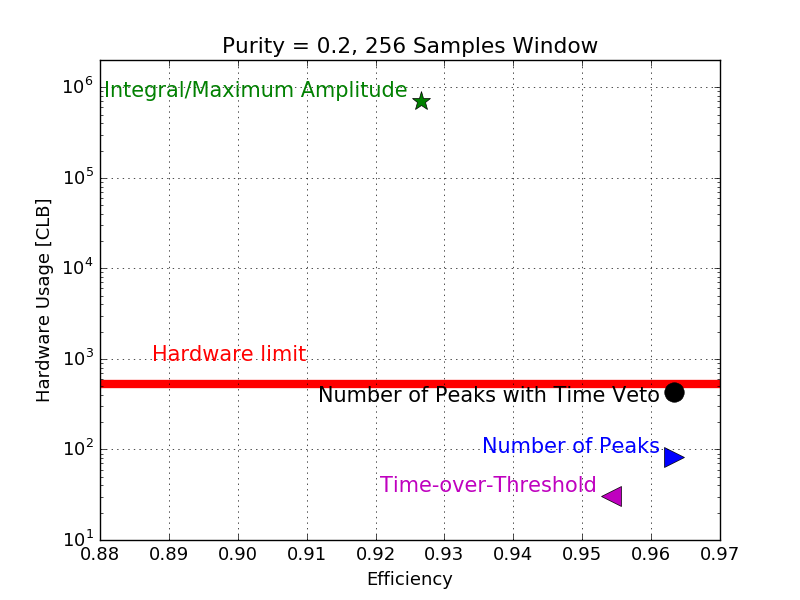}
\caption{Hardware usage of different feature extraction algorithms versus efficiency at a $20\%$ purity level and on the usual window size of $256$ samples (1 sample $\approx 20\mathrm{ns}$). The red line indicates hardware resource availability per channel.}
  \label{fig:usagevsefficiency}
\end{subfigure}
\caption{}
\label{fig:npeak}
\end{figure}

Many features have free parameters, such as the threshold $\theta$ in the Number-of-Peaks (Equation~\ref{eq:npeaksd2}) and Time-over-Threshold (Equation~\ref{eq:totd}) algorithms. These parameters have to be tuned for the optimum value. This can be achieved by sweeping the free parameter $\theta$ \cite{Eiben11parametertuning}, as shown in Figure~\ref{fig:tottuning}.  

\subsection{Correlation Analysis}
The features examined are correlated to a very high degree (Pearson's $r>0.8$), except the Integral-over-Amplitude that cannot be implemented, as shown in section~\ref{sec:implementation}. This indicates that redundant information would be obtained when using more than one feature. It can be concluded that no significant increase of performance can be expected when using more than one feature. Therefore a single threshold trigger on the feature output value is used, and other machine learning algorithms considered such as the perceptron and the feed-forward neural network have been discarded.
\subsection{Implementation}
\label{sec:implementation}
As 64 channel triggers plus the other firmware elements have to be synthesised on a single FPGA, hardware resources are strictly limited. Implementation of algorithms and their synthesis has been undertaken for the Time-over-Threshold, Number-of-Peaks (with and without time veto) and Integral-over-Amplitude features, as well as for the Integral feature. 

The Integral, Time-over-Threshold, Number-of-Peaks features rely on comparators, sums and differences only that can be easily implemented as adder-subtractors into digital logic \cite{nagaraj2012fpga}.
However, the Integral-over-Amplitude algorithm contains a division, an operation that is resource-intense on the FPGA. 
The usage of hardware resources is compared in Figure~\ref{fig:usagevsefficiency}. 
It can be seen that the Integral-over-Amplitude algorithm exceeds the hardware limit by far.  Therefore, implementation into the firmware has been made for the Number-of-Peaks algorithm  and the Time-over-Threshold feature extraction method.

\section{Conclusion}
According to the findings, the most suitable trigger has been implemented, that are now the Number-of-Peaks and the Time-over-Threshold feature extraction algorithms and a threshold trigger on the feature extraction output value. These algorithms have been implemented and synthesised into firmware, and embedded into the \textit{IPbus} framework. 
The trigger performance evaluation was carried out relative to the test data, and conditions at the reactor are not known at the moment. Therefore the decision on the trigger algorithm relies on the assumption that the best performing algorithms on the test environment also perform well on the reactor site. However this might be wrong if background signals are qualitatively different from the test signals. This can be considered unlikely though, as the same detection scheme underlies both setups. The actual background rate determines at which efficiency level the trigger can operate. The tuning of the parameters for the trigger is performed at \textbf{calibration}. Calibration will aim not only for highest efficiency, but also for uniformity along the different channels. As trigger efficiency directly propagates to detector efficiency, optimal trigger calibration is essential for the quality of the experiment.

The hunt for sterile neutrinos in the very-short baseline range has started, and SoLid is about to start its first large-scale physics run. In the case the reactor anomaly cannot be reconfirmed, SoLid's purpose might not be fulfilled, but the results would fit nicely not only with the three-flavour neutrino oscillation model, but also with recent results from accelerator- and atmospheric-based neutrino experiments. In case SoLid, in line with other very short-baseline experiments, does find a deficit, a theorist might hope for a systematic simulation error in calculating reactor neutrino flux to give sufficient explanation -- if not, he or she has to seriously consider an update of the neutrino model currently most favoured,  from having three to having four or even more neutrino flavour states by adding sterile neutrinos.

\Acknowledgements
I want to express gratitude to the collaborators of the SoLid read-out team, David Cussans, Dave Newbold, Steve Manley  and Nick Ryder.

\end{document}